\documentclass[]{spie}  
\usepackage[]{graphicx}

\title{New Method for Path-length Equalization of Long Single-mode Fibers for Interferometry} 

\author{M. Anderson\supit{a}, J.D. Monnier\supit{b},K.  Ozdowy\supit{b}, J. Woillez\supit{c}, G. Perrin\supit{d} \skiplinehalf
\supit{a}Georgia State University, 25 Park Place South, Suite 605, Atlanta, Georgia, USA; \\
\supit{b}University of Michigan, 500 Church St, Ann Arbor, Michigan, USA;\\
\supit{c}European Southern Observatory, Karl-Schwarzschild-Str. 2, Garching bei M{\"u}nchen 85748, Germany;\\
\supit{d}Observatoire de Paris, 61 avenue de l'Observatoire, 75014, Paris, France;\\
}

\authorinfo{Further author information: Matthew Anderson: E-mail: manderson@chara.gsu.edu}

\begin{document} 
  \maketitle 

\begin{abstract}
The ability to use single mode (SM) fibers for beam transport in optical interferometry offers practical advantages over conventional long vacuum pipes. One challenge facing fiber transport is maintaining constant differential path length in an environment where environmental thermal variations can lead to cm-level variations from day to night. We have fabricated three composite cables of length 470 m, each containing 4 copper wires and 3 SM fibers that operate at the astronomical H band (1500-1800 nm). Multiple fibers allow us to test performance of a circular core fiber (SMF28), a panda-style polarization-maintaining (PM) fiber, and a lastly a specialty dispersion-compensated PM fiber. We will present experimental results using precision electrical resistance measurements of the of a composite cable beam transport system. We find that the application of 1200 W over a 470 m cable causes the optical path difference in air to change by  75 mm (+/- 2 mm) and the resistance to change from 5.36 to 5.50 $\Omega$. Additionally, we show control of the dispersion of 470 m of fiber in a single polarization using white light interference fringes ($\lambda_c$=1575 nm, $\Delta$$\lambda$=75 nm) using our method.

\end{abstract}



\section{INTRODUCTION}

\label{sec:intro}  

Long baseline optical and infrared (OIR) interferometry attains the highest possible angular resolution for studying celestial objects by combining light from widely separated telescopes, beating the diffraction-limit of the individual apertures and boosting the resolution. One limit on the practical size of OIR interferometers is the loss associated with conventional beam transport through vacuum tubes via mirror trains. One proposed solution to this limit is the use of single mode (SM) fiber optics for beam transport instead of the vacuum tube optical trains. 

The astronomical H band (1500 - 1800 nm) in the near infrared (NIR) is well suited for use in SM fiber optics. The telecommunications industry has adopted the regime around 1.55  $\mu$m as the standard for optical communication because of the very low attenuation ($<0$.3 dB/km) in commonly available fibers. The lower light loss associated with fiber optics, at these wavelengths, will facilitate beam transport over distances of kilometers and directly address the technological problem recognized by the National Optical Astronomy Observatories (NOAO) workshop $``$New Direction for Interferometry$"$ \cite{NOAO}. Beyond allowing the development of much greater baselines, the use of fiber optics for beam transport will allow the development of sites where terrain limits the size and layout of an array of telescopes. Additional benefits of using fiber optics over conventional beam transport systems include reductions in both the infrastructure costs and environmental impact 

Substantial work pertaining to the deployment of SM fiber optics for beam transport has been carried out by the team working on the Optical Hawaiian Array for Nano-radian Astronomy (OHANA) \cite{2000SPIE.4006..708P}, including much of the theoretical groundwork required for organizing large optical interferomter arrays. The OHANA group has successfully used SM fiber optics to link the Keck 10-meter telescopes interferometrically \cite{2006Sci...311..194P}. 

Some of this work addresses the principle difficulty in using SM fiber optics for beam transport where temperature changes along different fiber links will induce variable differential dispersion and optical path differences (OPD) between the two optical paths. The OHANA group proposed compensating for the differential dispersion by two methods. First, the use of a $``$dispersive optical fiber delay line$"$ in which an additional length of fiber is stretched on a piezoelectric drum to compensate for the differential dispersion \cite{2004OptCo.232...31V}. Second, the use of bulk CaF2 with controlled variation of thickness is used \cite{2004OptCo.232...31V}.  In this paper, we discuss the development and initial testing of an additional method for compensating for the OPD and differential dispersion induced in the fiber optics of a beam transport system using precision temperature control of the fiber optic cable itself.

\section{Composite Cables Beam Transport} 

\subsection{Overview}

The composite cable beam transport system has been designed to address both OPD and dispersion effects induced by diurnal temperature fluctuations by controlling the temperature within the cable. In order to heat and measure the temperature of a cable hundreds of meters, or even kilometers, in length, the sensing and heating elements must either be very long, or numerous. For simplicity, in this system, the fiber optics are followed over their entire lengths by copper wires that act as both heating and sensing elements. The temperature of the fiber optic is then controlled via ohmic heating of the copper wire and the resistive properties of the copper are also used to measure the temperature. Containing the fiber optics, the heating element, and the sensing element within an insulated and well protected envelope renders the entirety of the system a single composite cable.

The concept for using ohmic heating to control the dispersion in fiber optics has already been shown and a great deal of the theoretical framework is discussed in Mueller, Ref. \citenum{Mueller:aa}. Though the method outlined in Ref. \citenum{Mueller:aa} is fundamentally similar to the system discussed in this paper, the system discussed here operates on much longer timescales and across the entire length of the fiber optic rather than a short portion of it.


\subsection{Construction} 

\label{sec:title}
  \begin{figure}
   \begin{center}
   \begin{tabular}{c}
   \includegraphics[height=7cm]{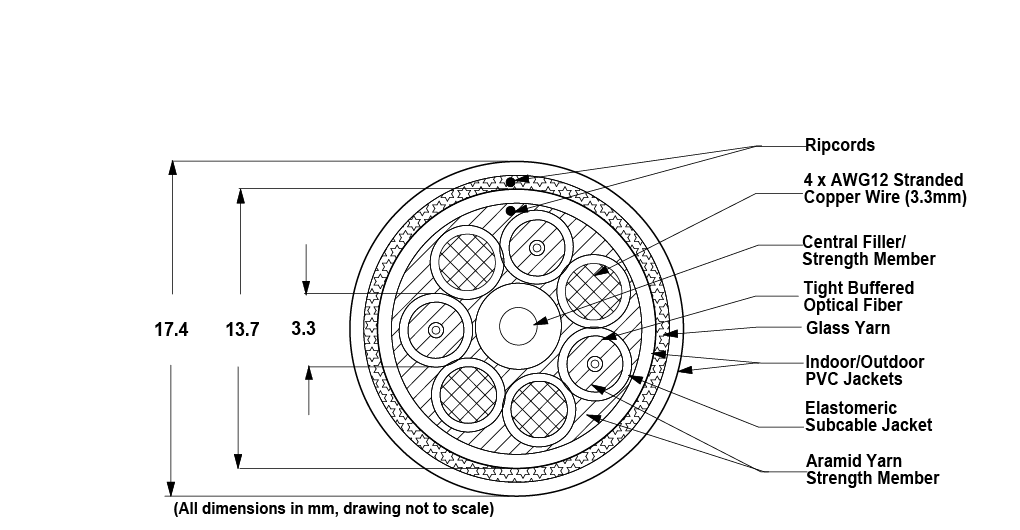}
   \end{tabular}
   \end{center}
   \caption[CableXSection] 
  { \label{fig:CableXSection} 
Cross-section of custom fiber/conductor composite cable assembly [PROPRIETARY]}
   \end{figure} 

The composite cables that comprise the heart of this beam transport system were manufactured by the Optical Cable Corporation (OCC) and many of the specific details of their construction remain proprietary.  The three 470 m long cables are built upon a preexisting commercial platform with only the fiber optics contained in the cable being different.
The cables contain 4 standard copper conductors (AWG12) rated to 600 V that are arranged along with the three fiber optic bundles around a central strengthening member, see figure \ref{fig:CableXSection}, and are connected at one end of the cable to form two independent loops through the length of the cable. This gives several possible configurations for heating and monitoring loops. The individual fiber bundles consist of the bare fibers as provided by their individual manufactures wrapped protective fibers and insulation. Each of the 4 copper conductors and the fiber optic enclosing bundles are wrapped around the central strengthening member along with the a insulation forming a helical structure. This whole bundle is then fully encased in two PVC wrappers separated by a final layer of strengthening fiber. According to OCC, the mechanics of constructing the cables introduce a maximum error in the lay length of 1$\%$ resulting in a worst-case scenario of 2$\%$ error across two of the composite cables. While the uneven lengths of the cables was corrected by cutting the final composite cables to the same overall length, in this case 470 m, the errors in the helical wrapping left the lengths of the fiber optics, optical path length (OPL), different between the cables. 

The differences in OPL within the cables made equalizing the path length to find initial fringes difficult. In order to find initial fringes with the test bed interferometer, the lengths of individual fibers had to be measured very precisely independently and trimmed to match. These measurements were made with a Luna Technologies optical backscatter reflectometer (OBR) 4600 and showed that the OPL of the fiber varied by up to 2 meters between the shortest and longest cables and up to 0.7 meters between the fibers in a single cable.  In reality, the helical wrapping error was far better than the maximum error specified by OCC, but still far from precise enough for interferometry.

\begin{figure}
   \begin{center}
   \begin{tabular}{c}
   \includegraphics[height=10cm]{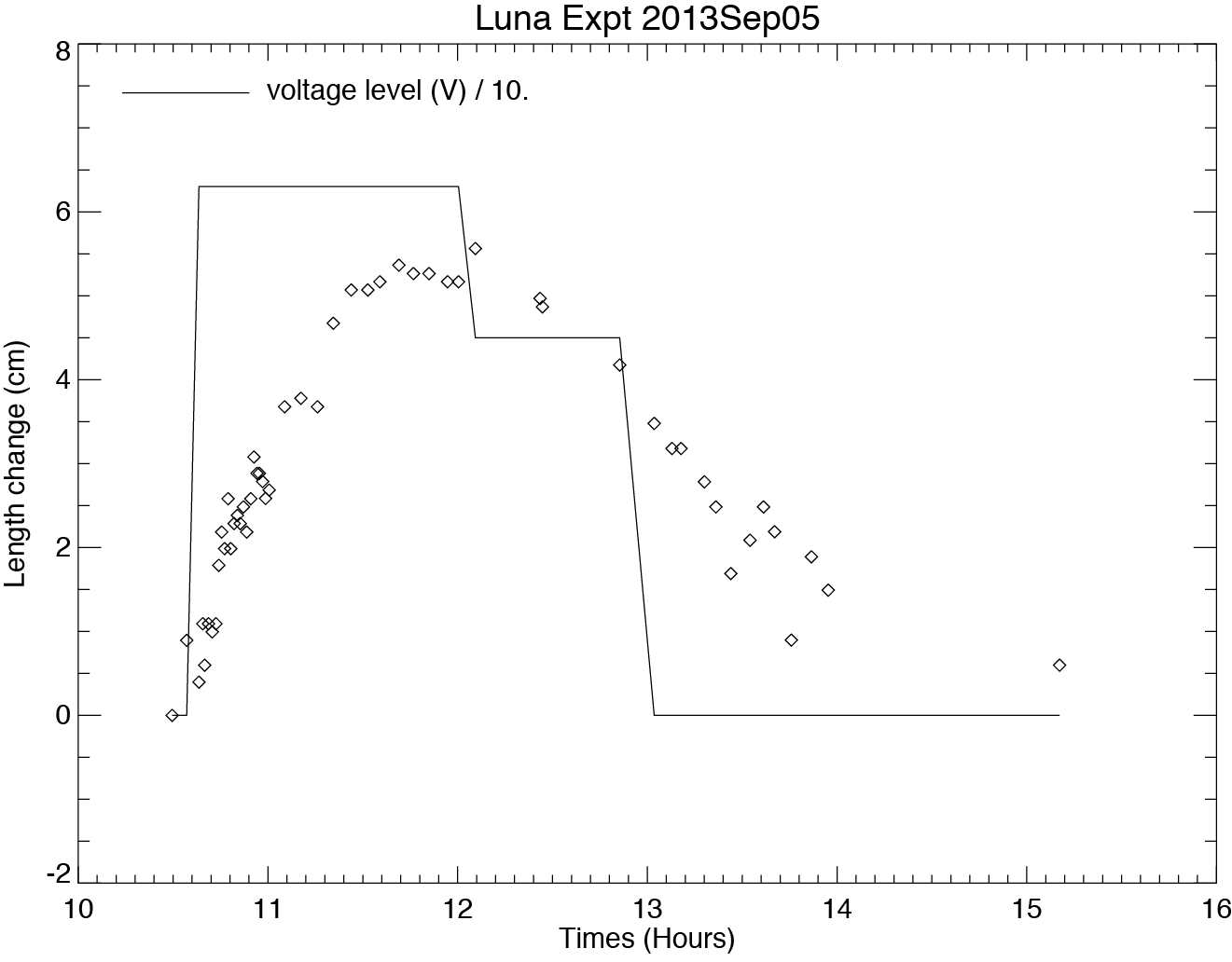}
   \end{tabular}
   \end{center}
   \caption[LunaTest] 
   { \label{fig:LunaTest} 
Test results from voltage testing of the Fiber Logix PM1550 fiber optic. The solid line represents the voltage set on the power supply and the diamonds represent length measurements made with a Luna Technologies OBR 4600.}
   \end{figure}

A basic test was conducted with the OBR where the length of the Fiber Logix PM1550 fiber optic was measured as 80 volts was applied to the cable. The results from one test are shown in figure \ref{fig:LunaTest}. With the application of 870 W, the length of the composite cable was measured to change by 5.5 cm. The figure shows that there is a delay in the time of the application of voltage and the equilibrium point of the cable length of approximately one hour.

\subsection{Fiber Optics} 
Three different fiber optics are included in the composite cables. One of them is a standard telecom fiber with a circular core, SMF 28e+, this particular fiber is included in the composite cable so that, if needed, a laser metrology system can be implemented to measure the length of the cables independently of the resistance feedback system. The remaining two fibers are intended as science fibers, both are polarization maintaing fibers. Polarization maintaining fiber optics are required for long distance beam transport as the birefringent effects of stresses and bending of standard fiber optic will degrade the fringe contrast.  \cite{2003SPIE.4838.1370K} One of the science fibers is a higher grade dispersion compensated fiber optimized for  $\lambda$=1550 nm. The two are included in the composite cable so that the performance of the two can be compared to each other without constructing additional cables. The dispersion compensated fiber is manufactured by Corning (PM DS 15-U40A) and should have dispersion characteristics on the order of 10x better than the other polarization maintaining fiber manufactured by Fiber Logix (PM1550 125-13/250C). Future testing will determine if the dispersion compensated fiber is required or worth the additional cost and attenuation, see table \ref{table:FiberSpecs}, or if the lower cost  Fiber Logix fiber is sufficient for beam transport.

\begin{table}[h]

\begin{center}       
\begin{tabular}{| l | c | c |} 
\hline
\rule[-1ex]{0pt}{3.5ex}  Fiber Optic &  Cutoff Wavelength $\mu$m & Attenuation (dB/km)  \\
\hline
\rule[-1ex]{0pt}{3.5ex}  Corning SMF28e+. & 1.260 & 0.200  \\
\hline
\rule[-1ex]{0pt}{3.5ex}  Corning PM DS 15-U40A & 1.073  & 0.277  \\
\hline
\rule[-1ex]{0pt}{3.5ex}  Fiber Logix PM1550 125-13/250C & 1.371 &  0.177 \\
\hline
\end{tabular}
\caption{Fiber optic specifications from manufacturer performance tests, the values are the mean of the three fibers of the specified type.} 
\label{tab:fonts}
\label{table:FiberSpecs}
\end{center}
\end{table}

\section{Electronics Control System}

\subsection{Power Supply}

   \begin{figure}
   \begin{center}
   \begin{tabular}{c}
   \includegraphics[height=10cm]{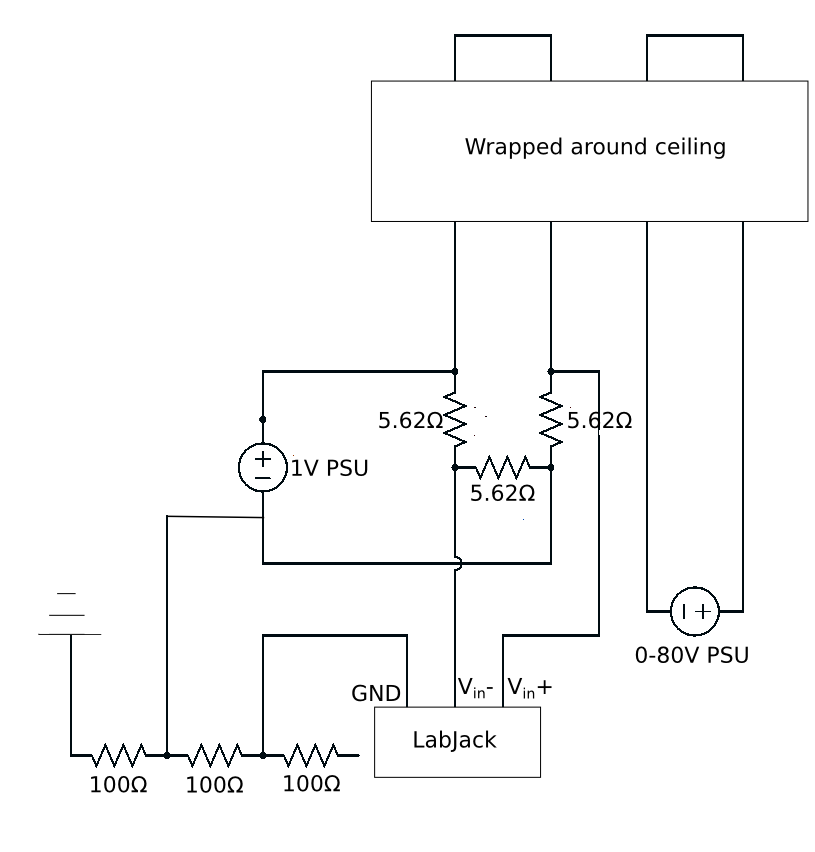}
   \end{tabular}
   \end{center}
   \caption[Bridge] 
   { \label{fig:Bridge} 
Wheatstone Bridge and voltage monitoring circuit used to monitor resistance and therefore length of the composite cable.   }
   \end{figure} 
Initially, the electronic heating system was designed around the use of Alternating Current (AC). A commercially available silicon controlled rectifying device was adopted for these early designs. These devices offered precise power control at approximately 33$\%$ the cost of conditioned direct current (DC) power supplies. Early testing revealed that the alternating current induced severe vibrations within the composite cable scrambling the optical signal in the fiber optic. The vibrations seen were are likely a result of the alternating current inducing forces between nearby current carrying wires. Due to this problem, the system was redesigned for DC control. The current power supply delivers variable power up to 1500 W.  While the system was designed for use of standard 120 VAC power to keep costs low and infrastructure requirements to a minimum, the rating of the conductors in the copper wire leaves the option for applying much higher power to the composite cable if required. 

\subsection{Resistance Monitoring} 
\label{subsec:resistance}
In order to control the length of the fiber optic, the temperature of the system is monitored using precise measurements of the resistance through the sensing loop. Using a simple model relating the length and resistivity of the copper conductors and the length and index of refraction of the fiber optic as a function of temperature that the resistance of the sensing loop would have to be controlled to a few milli-Ohms.
To achieve the desired resistance measurement precision, a Wheatstone Bridge is used. Three resisters of 5.62 $\Omega$, which roughly match the room temperature resistance (5.2 $\Omega$) of the cables are paired to a cable with the conductor loop as the fourth side of the bridge, see figure \ref{fig:Bridge}. The resisters purchased are off-the-shelf components with low temperature variance of +/- 15 ppm/$^\circ$C. The bridge is powered by a very stable low voltage power supply. Resistance is measured by measuring the voltage drop across the bridge using a high precision (22-bit effective resolution) digital to analog converter (DAC). 

\subsection{Proportional-Integral-Derivative (PID) Control}

Through the University of Michigan's undergraduate research opportunity program (UROP), a program that links freshmen and sophomores with active research projects on campus, a student came to work in the laboratory on the electronic control for the composite cable system. Control of the composite cables will be accomplished through a PID controller that will monitor the resistance in the sensing loop and keep this value stable at some desired set point.

\begin{figure}
   \begin{center}
   \begin{tabular}{c}
   \includegraphics[height=10cm]{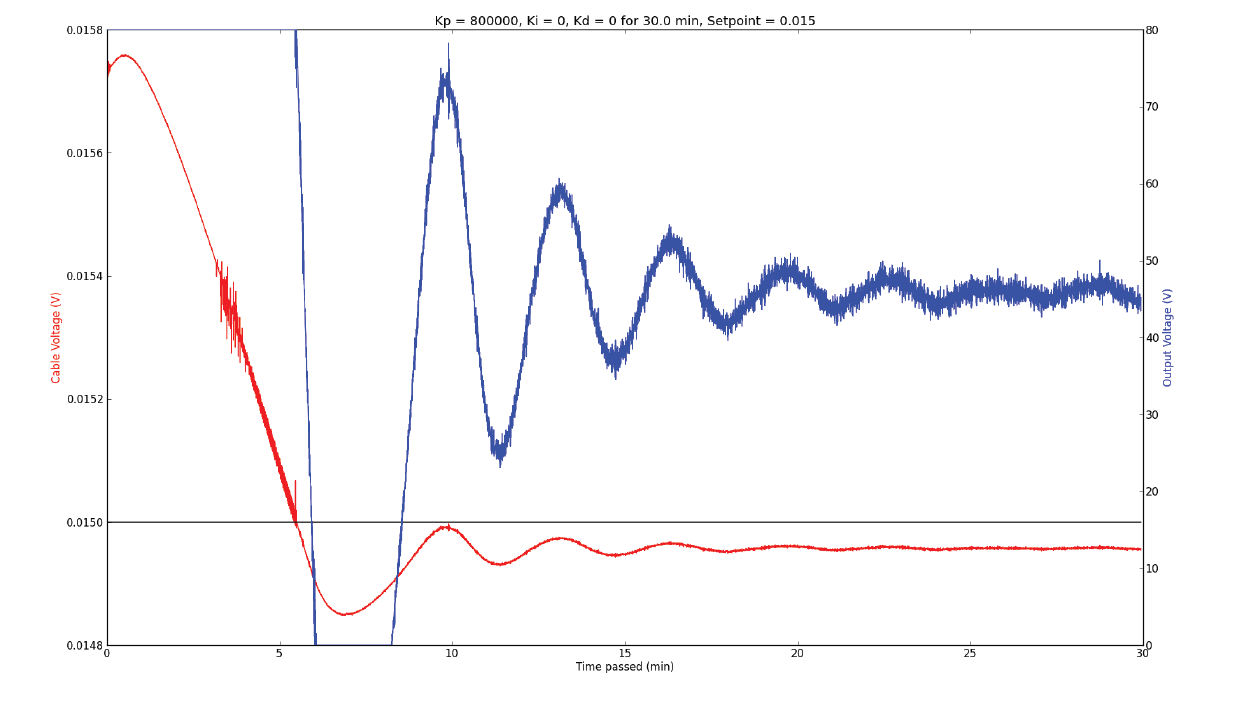}
   \end{tabular}
   \end{center}
   \caption[PIControl] 
   { \label{fig:PIControl} 
Results of proportional-integral control of the bridge voltage (resistance). Output voltage represents the set voltage for the DC power supply and Cable Voltage represents the voltage measured across the Wheatstone bridge. }
   \end{figure}

To date, successful implementation of the proportional and integral portions of the PID loop have been accomplished.  Current performance of the PI control loop can be seen in figure \ref{fig:PIControl}. The approximately 20 minute delay in stabilization and  0.05 m$\Omega$ offset from the set point are expected to improve with further refinement and addition of the derivative term. Once fully implemented, PID loops on both composite cable lengths should maintain the OPD at a position of minimum dispersion.

\section{Interferometric Testing}

   \begin{figure}
   \begin{center}
   \begin{tabular}{c}
   \includegraphics[height=6cm]{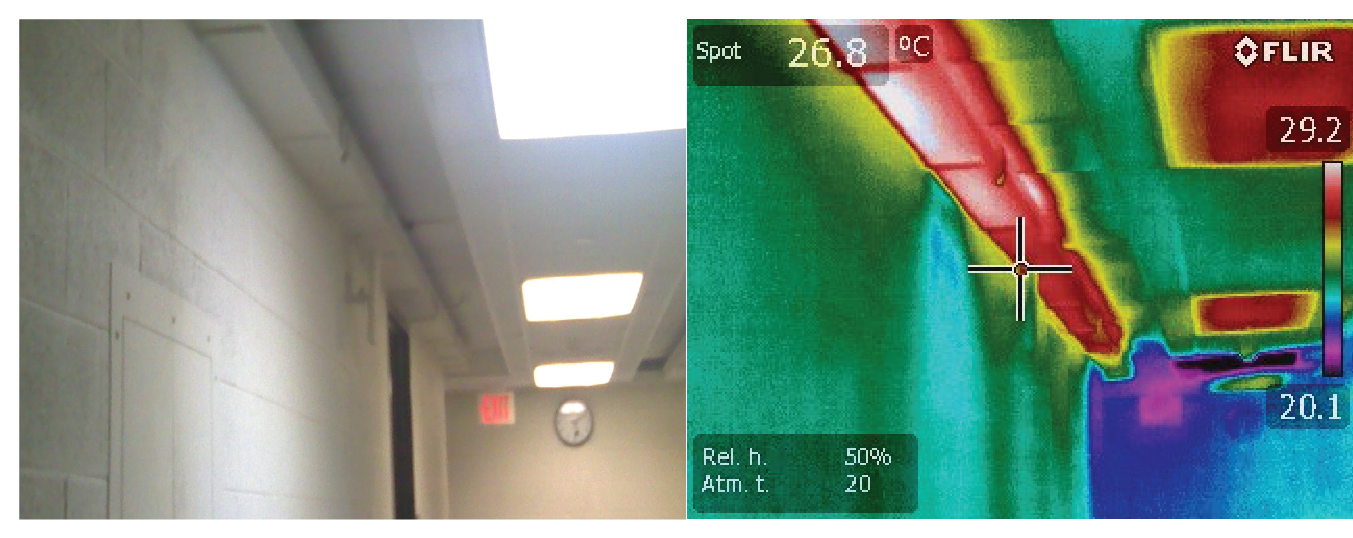}
   \end{tabular}
   \end{center}
   \caption[Ceiling] 
   { \label{fig:Ceiling} 
The cable tray around the inside of the laboratory building contains five coils of the control cable. The temperature of the tray is heated approximately 5$^\circ$C above the surrounding temperature by the powered cable. }
   \end{figure} 




   \begin{figure}
   \begin{center}
   \begin{tabular}{c}
   \includegraphics[height=8cm]{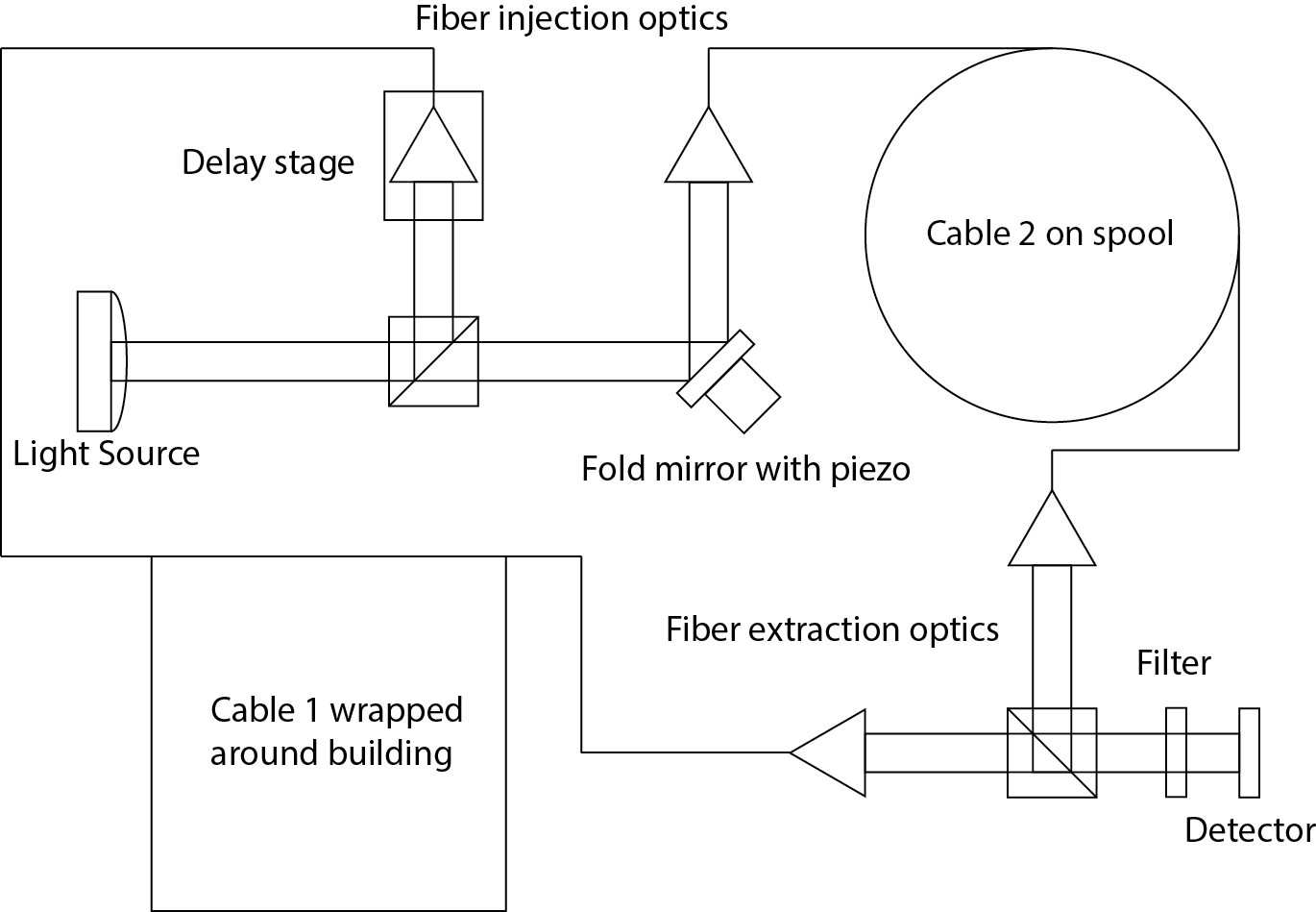}
   \end{tabular}
   \end{center}
   \caption[OpticsLayout] 
   { \label{fig:OpticsLayout} 
Optical layout of the interferometer used to test temperature control of the composite cable.   }
   \end{figure} 

\subsection{Layout}
To test the viability of using the composite cables an interferometer was required. Using components already in place at the University of Michigan as well as a large number of new components, a testbed interferometer was built to facilitate testing the cables. In order to create temperature differences across the two cables, one arm of the testbed interferometer passed through a composite cable on a spool within the laboratory, the second arm passed through another cable located in a cable tray wrapped around the inside of the building, see figure \ref{fig:Ceiling}. This configuration put the two cables into different environments and allowed the arm outside of the lab to be heated without concern for the self insulating effects seen while the cable was still wound on a spool.

The fiber optics of the composite cables are fed with a fiber-coupled white light source that is linearly polarized and then split by a beamsplitter before the injection optics. The injection optics for the controlled cable are mounted on a 60 mm translation stage that acts as a small delay line. The uncontrolled cable's input path is modulated by a 300 $\mu$m stroke piezo. The light passes through the two composite cables before being recombined through another beam splitter, see fig. \ref{fig:OpticsLayout}. Both sets of extraction optics are mounted along hard-stop rails which allow the coarse adjustment of OPD. An infrared photodiode is used as a detector with a narrow band filter centered at 1575 nm with a full width at half maximum (FWHM) of 76 nm.

\subsection{Testing Procedure and Results}

\begin{figure}
   \begin{center}
   \begin{tabular}{c}
   \includegraphics[height=6cm]{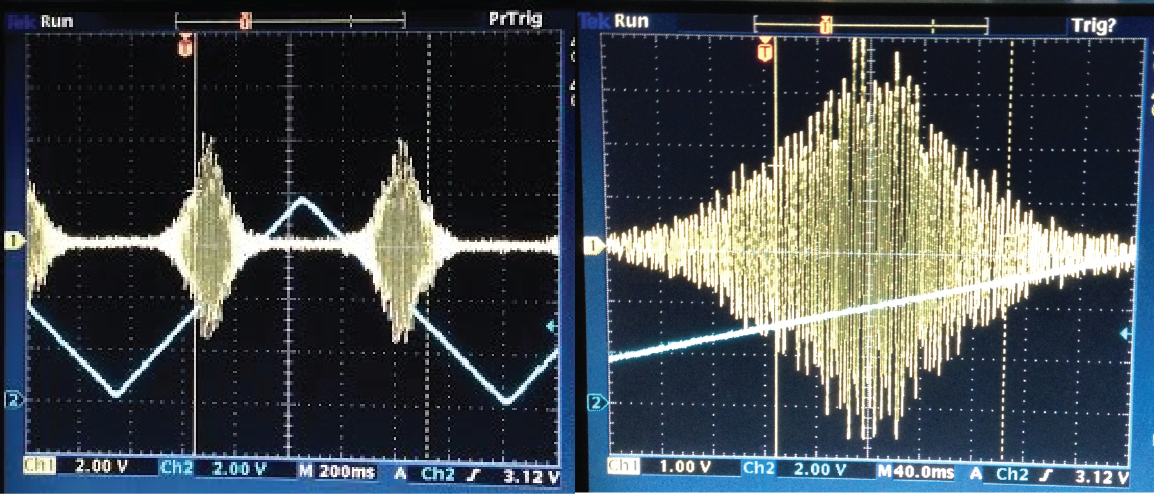}
   \end{tabular}
   \end{center}
   \caption[IRFringes] 
   { \label{fig:IRFringes} 
White light fringes found in interferometric testing of the composite cables at different applied voltages.  }
   \end{figure}

\begin{figure}
   \begin{center}
   \begin{tabular}{c}
   \includegraphics[height=8cm]{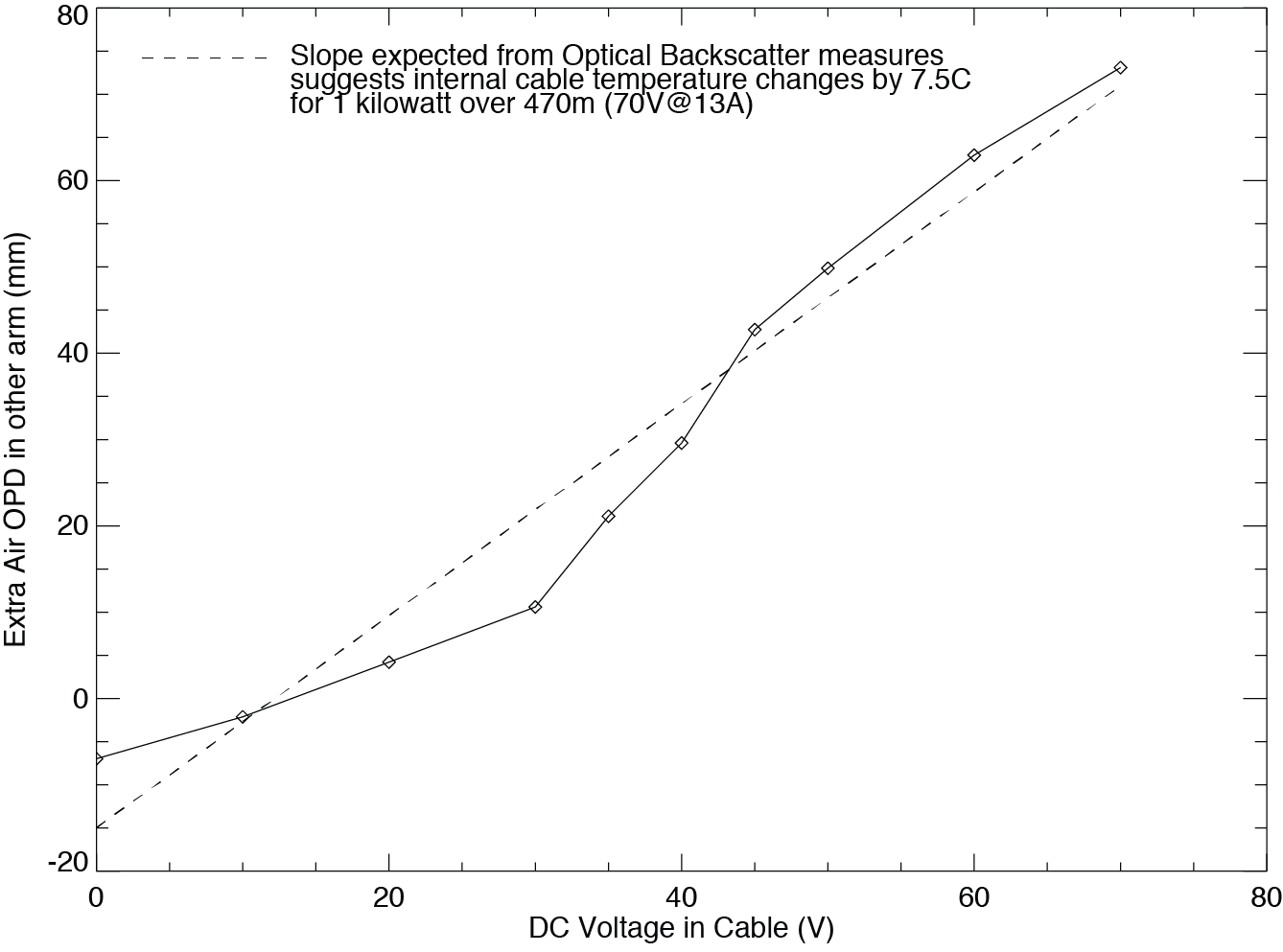}
   \end{tabular}
   \end{center}
   \caption[VoltsVSOPD] 
   { \label{fig:VoltsVSOPD} 
Voltage testing data shows a reasonable correlation to the expected slope derived from the results of the OBR device tests.  }
   \end{figure}

 Once the first fringes with a white light source were found, figure \ref{fig:IRFringes}, testing of the composite cable system began. To test the dispersive behavior of the cable over the range of temperatures available with the power supply, the maximum voltage was applied and the cable was allowed to equilibrate. Then, the voltage was dropped by 10 volt increments and allowed to settle before data was taken. Around what appeared to be the best dispersion conditions, estimated by eye and the appearance of side lobes to the fringe packets, the voltage was incremented by 5 volts. OPD was measured using the position of the delay line with additional adjustments made to the interferometer corrected. Figure \ref{fig:VoltsVSOPD} shows the OPD measurement as a function of the set voltage. While the data are in agreement with the expected behavior from the OBR experimental data discussed previously, the correlation is somewhat rough. The set voltage does not take into account the actual temperature of the cable, so it was not expected to precisely follow the trend found with the OBR. 
 
\begin{figure}
   \begin{center}
   \begin{tabular}{c}
   \includegraphics[height=8cm]{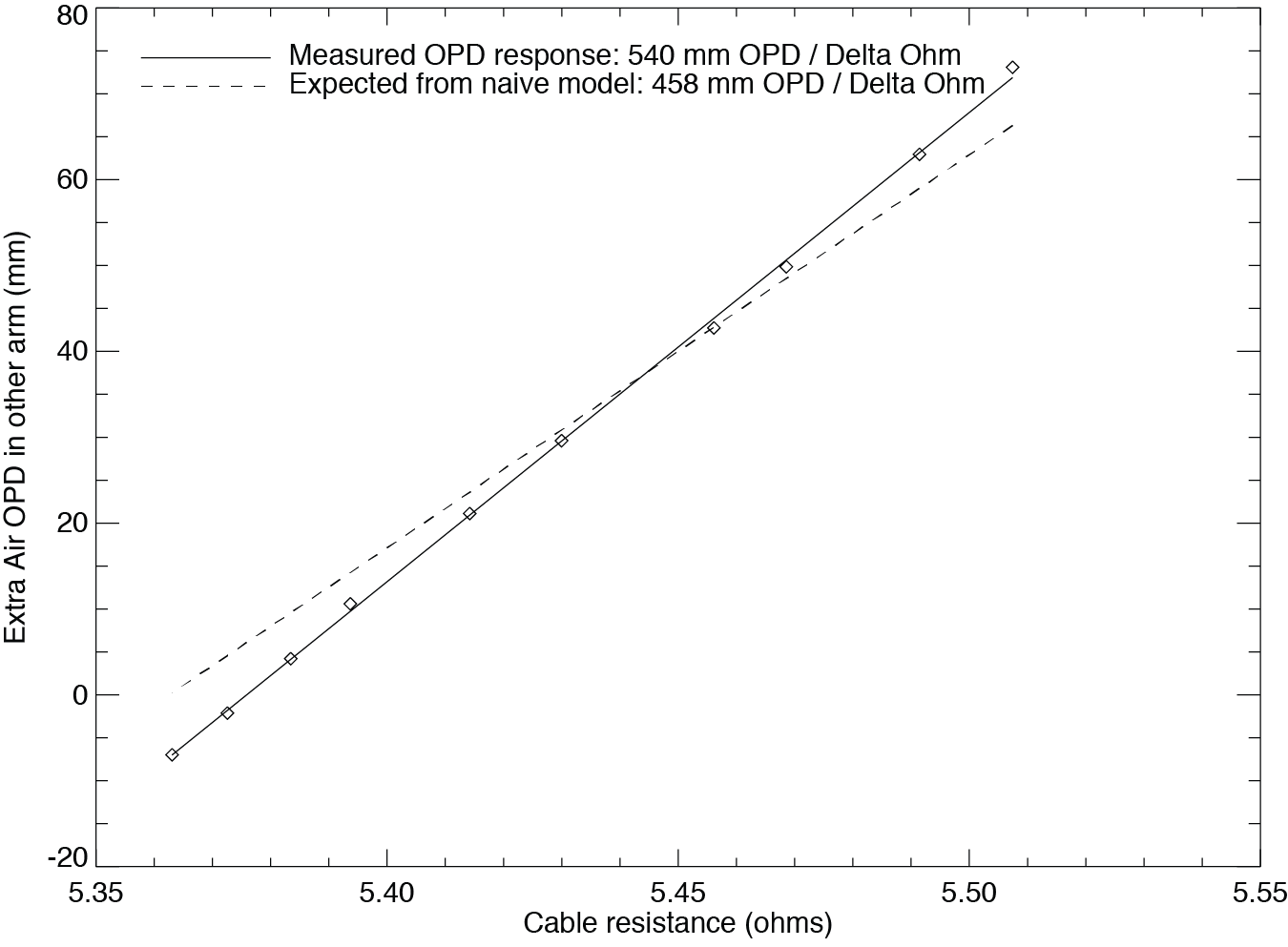}
   \end{tabular}
   \end{center}
   \caption[ResiVSOPD] 
   { \label{fig:ResiVSOPD} 
Comparing the OPD to the resistance of the cable measured across the bridge, a better measure for OPD very closely matches the simple model used initially.  }
   \end{figure}

Since the resistance across the loop of copper within the cable is a measurement of the temperature of the cable, it was expected to more accurately track with the OPD and dispersion within the cable than the set voltage does. In figure \ref{fig:ResiVSOPD} the OPD is plotted as a function of the resistance across the cable rather than the set voltage. Over the range of power applied, from 0 to 1200 W, the OPD in air changed by approximately 75 mm (+/- 2 mm) and the resistance of the sensing loop changed by 0.14 $\Omega$. Since the environmental changes, namely temperature, that change the OPD and dispersion in the fiber optic change the resistance of the cable, the response is a far smoother function as expected. The data from the test measurements are plotted along with the line for model discussed in section \ref{subsec:resistance}. 

\begin{figure}
   \begin{center}
   \begin{tabular}{l}
   \includegraphics[height=10cm]{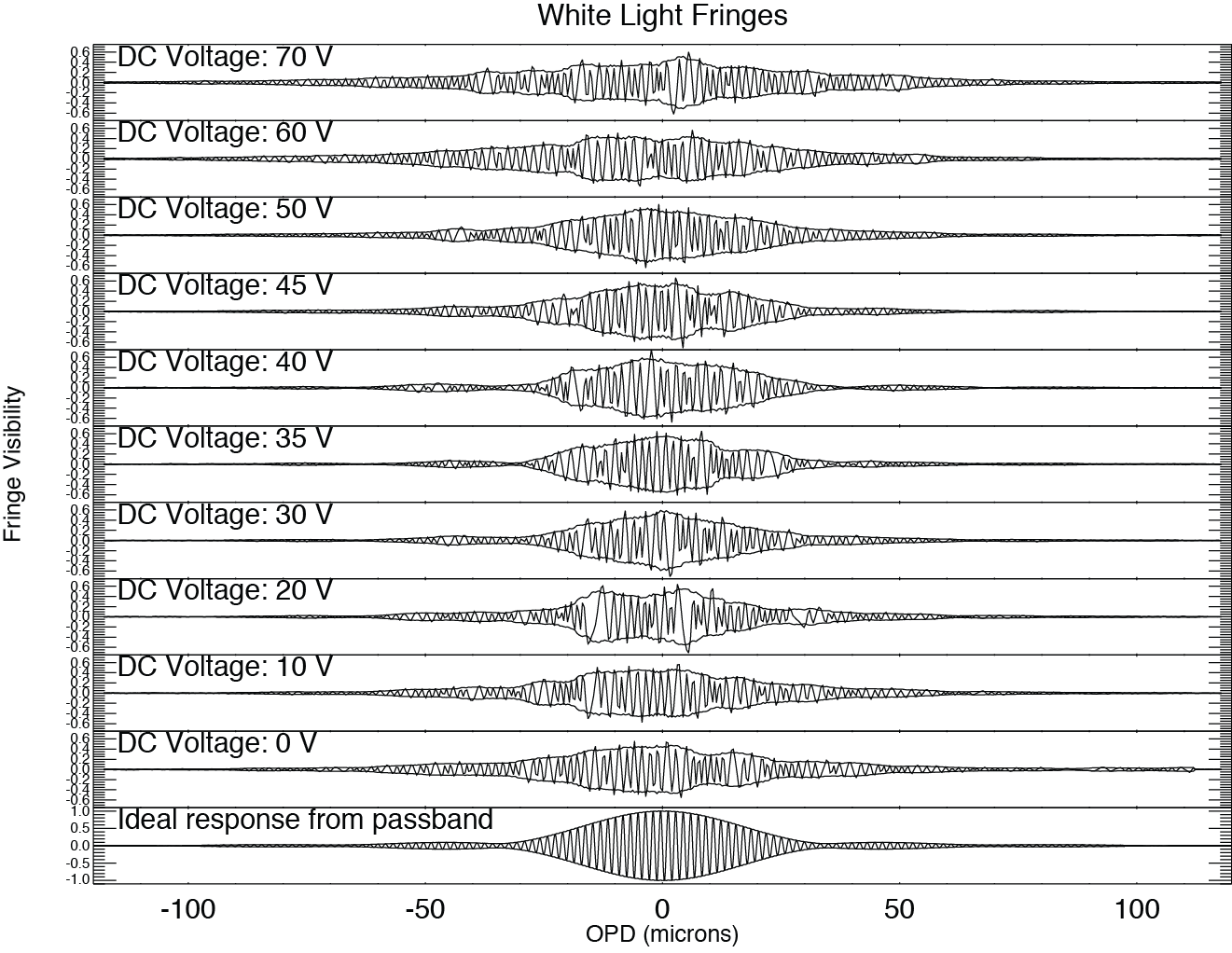}
   \end{tabular}
   \end{center}
   \caption[WLFringes] 
   { \label{fig:WLFringes} 
White light fringes with true visibilities taken at the various voltage settings applied. Visibilities suffer from the poor alignment of the backend of the interferometer. The bottom pane shows the ideal response from the narrow-band filter.  }
   \end{figure}

To begin to analyze the dispersive behavior of the fiber optics as the temperature changed the wave packets were fitted with Gaussian profiles and the full width at half maximum (FWHM) was taken. The FWHM is a direct analog to dispersion in the fiber. The wave packets for the voltage settings used ranging from 80 volts to 0 volts are shown in figure \ref{fig:WLFringes}. These wave packets are plotted at their true visibilities and careful examination of the width of the fringe packets indicates that we do indeed pass through a minimum in dispersion. The visibilites here reach 70$\%$ of the system visibility with minimal optimization. We believe that higher contrast will be possible with further alignment.

\begin{figure}
   \begin{center}
   \begin{tabular}{c}
   \includegraphics[height=10cm]{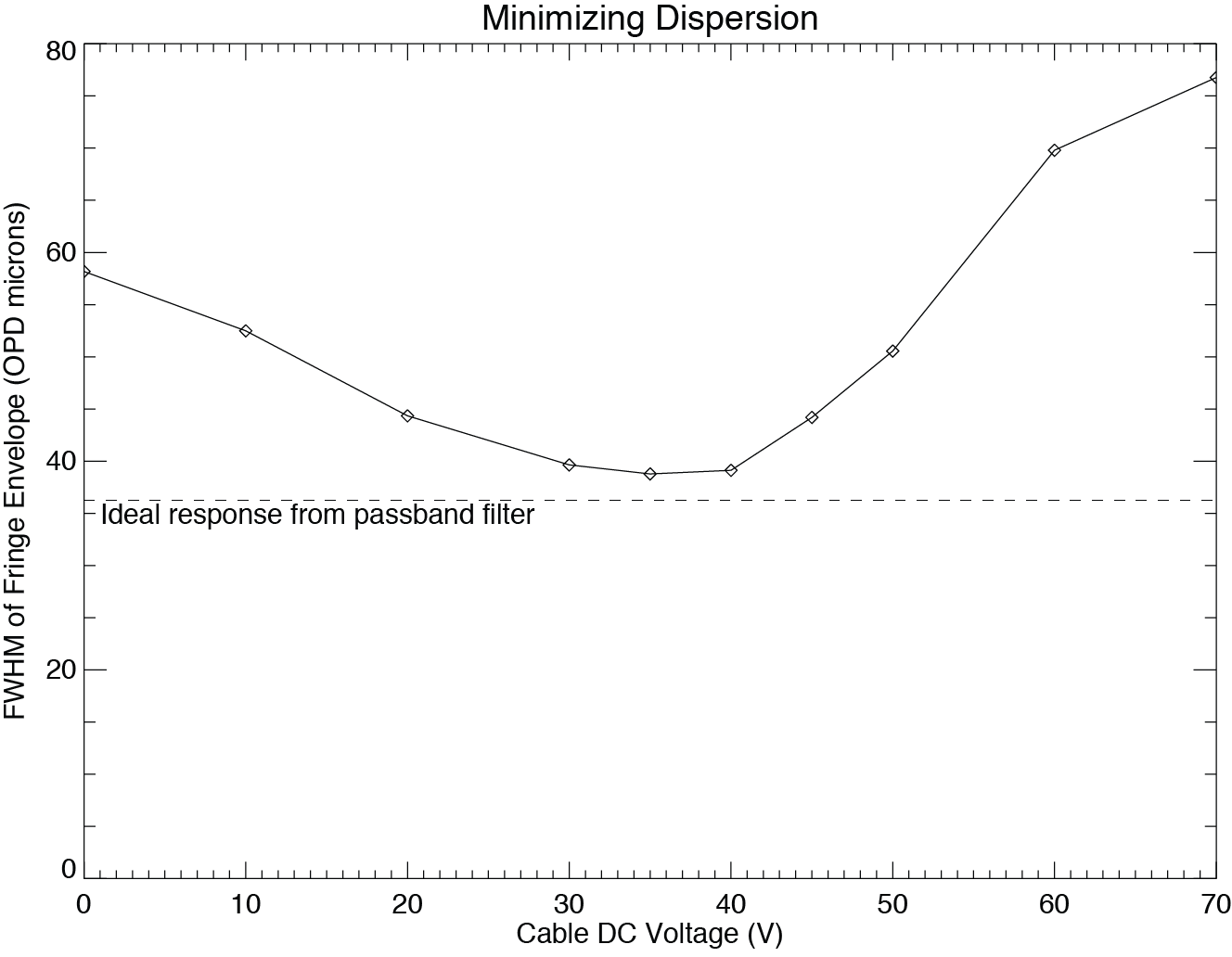}
   \end{tabular}
   \end{center}
   \caption[Dispersion] 
   { \label{fig:Dispersion} 
Dispersion as a function of the voltage setting. The full width at half maximum (FWHM) of the wave packet envelopes fitted with a gaussian profile are used as an analog of the dispersion. }
   \end{figure}

The FWHM of the fringe packets are plotted in figure \ref{fig:Dispersion} as a function of the set voltage. The ideal FWHM is plotted as a horizontal line, computed from the characteristics of the filter. At the minimum measured FWHM, around 35 V, the measurements approach the ideal response. In all, this indicates that the system is indeed working within the confines of the relatively quiescent building environment.


\section{Conclusions}
In this paper, we have discussed the development of a composite cable system for thermally controlled fiber optic beam transport for optical interferometry. We have shown white light fringes through two 470 m lengths of polarization maintaing fiber optics. We have also demonstrated basic control of the dispersive behavior of light through one 470 m composite cable as an arm of a testbed interferometer.

In the future, further testing of the composite cable system will be conducted. Planned tests include, comparisons between the two types of polarization maintaing fiber contained in the composite cable, closed-loop control of both composite cable arms of the testbed interferometer, and possibly on sky testing at the CHARA array.

\section{Acknowledgments} 
We acknowledge useful discussions with Mike Ireland and David Monnier. We also recognize the University of Michigan for funding initial work on this project through the LSA-APS fund and the UROP program. We also note our great appreciation for the assistance of OCC in working with us on this project.





\bibliography{spie}   

@ARTICLE{Mueller:aa,
	Title = {Optical fibers with interferometric path length stability by controlled heating for transmission of optical signals and as components in frequency standards},
	Author = {Holger Mueller and Achim Peters and Claus Braxmaier},
	Eprint = {physics/0511072},
	Month = {April},
	Url = {http://arxiv.org/abs/physics/0511072},
	Year = {2006}
}

@INPROCEEDINGS{2000SPIE.4006..708P,
   author = {{Perrin}, G. and {Lai}, O. and {Lena}, P.~J. and {Coud{\'e} du Foresto}, V.},
    title = "{Fibered large interferometer on top of Mauna Kea: OHANA, the optical Hawaiian array for nanoradian astronomy}",
	booktitle = {Interferometry in Optical Astronomy},
    year = 2000,
   	series = {Society of Photo-Optical Instrumentation Engineers (SPIE) Conference Series},
   	volume = 4006,
   	editor = {{L{\'e}na}, P. and {Quirrenbach}, A.},
    month = jul,
    pages = {708-714},
   	adsurl = {http://adsabs.harvard.edu/abs/2000SPIE.4006..708P}
}

@ARTICLE{2004OptCo.232...31V,
   author = {{Vergnole}, S. and {Delage}, L. and {Reynaud}, F.},
    title = "{Accurate measurements of differential chromatic dispersion and contrasts in an hectometric silica fibre interferometer in the frame of 'OHANA project}",
  journal = {Optics Communications},
     year = 2004,
    month = mar,
   volume = 232,
    pages = {31-43},
      doi = {10.1016/j.optcom.2003.12.052},
   adsurl = {http://adsabs.harvard.edu/abs/2004OptCo.232...31V}
}

@INPROCEEDINGS{NOAO,
   author = {NOAO},
	title = "{Workshop on the Future Directions for Ground-based Optical Interferometry}",
	booktitle = {},
    year = 2007,
   	series = {Society of Photo-Optical Instrumentation Engineers (SPIE) Conference Series},
   	month = apr,
    pages = {},
   	url = {http://www.noao.edu/meetings/interferometry/Workshop-report.pdf},
}

@ARTICLE{2006Sci...311..194P,
   author = {{Perrin}, G. and {Woillez}, J. and {Lai}, O. and {Gu{\'e}rin}, J. and 
	{Kotani}, T. and {Wizinowich}, P.~L. and {Le Mignant}, D. and 
	{Hrynevych}, M. and {Gathright}, J. and {L{\'e}na}, P. and {Chaffee}, F. and 
	{Vergnole}, S. and {Delage}, L. and {Reynaud}, F. and {Adamson}, A.~J. and 
	{Berthod}, C. and {Brient}, B. and {Collin}, C. and {Cr{\'e}tenet}, J. and 
	{Dauny}, F. and {Del{\'e}glise}, C. and {F{\'e}dou}, P. and 
	{Goeltzenlichter}, T. and {Guyon}, O. and {Hulin}, R. and {Marlot}, C. and 
	{Marteaud}, M. and {Melse}, B.-T. and {Nishikawa}, J. and {Reess}, J.-M. and 
	{Ridgway}, S.~T. and {Rigaut}, F. and {Roth}, K. and {Tokunaga}, A.~T. and 
	{Ziegler}, D.},
    title = "{Interferometric coupling of the Keck telescopes with single-mode fibers}",
  journal = {Science},
     year = 2006,
    month = jan,
   volume = 311,
    pages = {194},
      doi = {10.1126/science.1120249},
   adsurl = {http://adsabs.harvard.edu/abs/2006Sci...311..194P}
}

@INPROCEEDINGS{2003SPIE.4838.1370K,
   author = {{Kotani}, T. and {Nishikawa}, J. and {Sato}, K. and {Yoshizawa}, M. and 
	{Ohishi}, N. and {Fukushima}, T. and {Torii}, Y. and {Matsuda}, K. and 
	{Kubo}, K. and {Iwashita}, H. and {Suzuki}, S.},
    title = "{Long-baseline optical fiber interferometer instruments and science}",
booktitle = {Interferometry for Optical Astronomy II},
     year = 2003,
   series = {Society of Photo-Optical Instrumentation Engineers (SPIE) Conference Series},
   volume = 4838,
   editor = {{Traub}, W.~A.},
    month = feb,
    pages = {1370-1377},
   adsurl = {http://adsabs.harvard.edu/abs/2003SPIE.4838.1370K}
}
\bibliographystyle{spiebib}   

\end{document}